\documentclass{aa}

\def\Mso{{M$_{\rm \odot}$}}

\usepackage{txfonts}
\usepackage{verbatim}
\usepackage{graphicx}
\usepackage{longtable,lscape}
\usepackage{supertabular}
\usepackage{threeparttable}
\usepackage{natbib}
\usepackage{caption}
\pdfoutput=1
\bibpunct{(}{)}{;}{a}{}{,}

\begin{document}

\title{Compact planetary nebulae in the Galactic disk: analysis of the central stars\thanks{Based on observations made with the NASA/ESA \textit{Hubble Space Telescope}, obtained at the Space Telescope Science Institute, which is operated by the Association of Universities for Research in Astronomy, Inc., under NASA contract NAS 5--26555}}

\author{Manuel Moreno-Ib\'a\~nez \inst{1,2,3}
\and Eva Villaver \inst{2}
\and Richard A. Shaw \inst{4} 
\and Letizia Stanghellini \inst{4}}

\institute{Institute of Space Sciences (IEEC-CSIC), Meteorites, Minor Bodies and Planetary Science Group, Campus UAB, Carrer de Can Magrans, s/n E-08193 Cerdanyola del Vall\'es, Barcelona, Spain; \textit{mmoreno@ice.csic.es}
\and Universidad Aut{\'o}noma de Madrid, Departamento de
  F{\'i}sica Te{\'o}rica, Cantoblanco, E-28049 Madrid, Spain; \textit{eva.villaver@uam.es}
\and Finnish Geospatial Research Institute, Department of Geodesy and Geodynamics, Geodeetinrinne 2, FI-02431 Masala, Finland
\and National Optical Astronomy Observatory, 950 N.\ Cherry Avenue,
Tucson, AZ 85719, USA}

\abstract
{ 
We have obtained multi-wavelength observations of compact Galactic planetary nebulae (PNe) to probe post-asymptotic giant branch (AGB) evolution from the onset of nebular ejection. 
Here we analyze new observations from \textit{HST} to derive the masses and evolutionary status of their central stars (CSs).
} 
{ 
Our objective here is to derive the masses of the CSs hosted by compact PNe in order to better understand the relationship between the CS properties and those of the surrounding nebulae. 
We also compare this sample with others we obtained using the same technique in different metallicity environments: the Large and Small Magellanic Clouds. 
}
{
This paper is based on \textit{HST}/WFC3 images of 51 targets obtained in a snapshot survey (GO--11657). 
The high spatial resolution of \textit{HST} allows us to resolve these compact PNe and distinguish the CS emission from that of their surrounding PNe. 
We derive CS bolometric luminosities and effective temperatures using the Zanstra technique, from a combination of \textit{HST} photometry and ground-based spectroscopic data. 
The targets were imaged through the filters F200LP, F350LP, and F814W from which we derive Johnson $V$ and $I$ magnitudes. 
We infer CS masses by placing the stars on a temperature-luminosity diagram and compare their location with the best available, single star post-AGB evolutionary tracks.
}
{
We present new, unique photometric measurements of 50 CSs, and we derive effective temperatures and luminosities for most of them. 
Central star masses for 23 targets were derived with the evolutionary track technique; the remaining masses were indeterminate most likely because of underestimates of the stellar temperature, or because of substantial errors in the adopted statistical distances to these objects. 
We expect these problems will be largely overcome when the GAIA distance catalog becomes available. 
We find that objects with the higher ratios of Zanstra temperatures T(\ion{H}{i})/T( \ion{He}{ii} ) tend to have lower-mass progenitors. 
}
{
The distribution of CS masses in this sample of compact PNe is remarkably different from samples in the LMC and SMC, but with a median mass of 0.59~\Mso it is similar to other Galatic samples. 
Finally, we conclude from the typically advanced evolutionary state of the CSs on the log~$L$, log~$T_{eff}$ plane that the compact nature of many of the PNe is a result of their large distance, rather than their physical dimension.
}
\keywords{Galaxy--planetary nebulae: general--stars: AGB and post-AGB--stars: evolution--stars: fundamental parameters
 }

\maketitle

\section{Introduction}
Central stars (CSs) of planetary nebulae (PNe) are the end products of the  evolution of low- and intermediate-mass stars (between 1 and up to 8 
\Mso), the stars that most likely go through the asymptotic giant branch (AGB) phase. 
By the time a star departs the AGB it has lost most of its outer envelope, and if the remnant core evolves to high temperature before the ejected envelope disperses it will be visible as a PN for a brief ($\sim$10,000~yr) period. 
During the post-AGB phase the system evolves into two separate yet interconnected remnants: the gaseous, ionized part (PN), and the CS which later becomes a white dwarf. 
The evolution of the star after its departure from the AGB phase is mostly dependent on its mass \citep{VW1993, Wachter2002}. 
The CS mass is the primary determinant of the instantaneous rate and integrated amount of energy injected into the nebular shells, and thus it constraints the evolutionary timescale of the PN \citep{Vmg:02,Peri}. 
Since the determination of CS masses constrains fundamental empirical quantities such as the initial--final mass relation and dredge-up efficiency on the AGB, an accurate determination of CS masses for specific PN samples is very important to understanding post-AGB evolution, which has important applications in astrophysics. 

Because PNe are bright in narrow emission lines they are easy to discover, but their short lifetimes make them intrinsically rare. 
Thus the typical distance to a Galactic PN is large, often measured in kpc, and therefore few PN distances have been determined using direct techniques such as parallax. 
Indeed, accurate distances to Galactic PNe are known only for a handful of objects \citep{SSV2008} out of a few thousand catalogued (see e.g., \citealt{Par06}). 
As a result statistical distance methods have been developed and used over the years, which invariably results in large uncertainties \citep[see e.g.,][]{SSV2008}. 
This well known distance problem propagates to the determination of CSs masses which also suffer significant uncertainties (see, e.g., \citealt{SK85,SK1989,Setal:00}), although there are other limitations as well. 
In addition, the surface brightness of the nebular continuum can be very high compared to the CS continuum \citep{SK85}. 
This problem can be especially acute for angularly small PNe, where the CS cannot be spatially resolved from the surrounding nebula.

The problem with distance uncertainties in the determination of CS masses has been alleviated by observing PNe in the Magellanic Clouds with the \textit{Hubble Space Telescope} ({\it HST}) \citep{VSS2003, VSS:04,VSS:07}, where the distance is known to $\sim$10\% \citep{Ben:02}, and the distances to all nebulae within a Cloud are identical to within a few percent. 
Analyzing the combined samples of stars in the Small and Large Magellanic Clouds, \cite{VSS:07} found a higher average mass for the LMC CSs than for both white dwarfs and CSs of Galactic PNe \citep{SVMG2002}. 
This result was interpreted as a new way of probing the metallicity dependence on mass-loss rates during the AGB, where mass-loss at higher metallicity (i.e., in the Galactic population) is much more efficient at removing the AGB envelope than at lower metallicity (i.e., in the Magellanic Clouds).

Studies of PNe in the Galactic disk and elsewhere typically do not include angularly small targets. 
Such angularly small objects are sometimes supposed to be physically compact as well, suggesting a young nebular age. 
If the star cannot be spatially distinguished from the nebula, then the CS temperature and luminosity must be determined using photoionization modeling of the nebula without the important constraint of a CS continuum magnitude, as was done in the Magellanic Cloud before \textit{HST} was available (see e.g., \citealt{BL,Hlb:89}). 
As a result, and for the most part when the nebula cannot be distinguished from the CS, these studies can only determine lower limits to the stellar luminosities (see e.g., \citealt{VSS2003}). 

In this paper we analyze the CS properties of a sample of angularly compact Galactic PNe. 
The PNe have been observed with the {\it HST} in a snapshot survey (GO~11657) using the Wide-Field Camera~3 \citep[WFC3;][]{Kimble_etal08}, and the acquired images can be used both for nebular and stellar studies, as PNe and CSs are spatially distinguished in most cases. 
The images of the PNe, together with their morphological characteristics, were presented by  \cite{SSV16}(hereafter SSV16). 
In Sect. Sect. \ref{section2} we summarize the observations of the CSs and the photometric analysis, in section Sect. \ref{section3} we present the derived physical parameters of the selected stars. 
The inferred CS masses are presented in Sect. \ref{section4} together with a general discussion of our results; the conclusions are presented in Sect. \ref{section5}.

\section{Observations, sample description and analysis} \label{section2}

The goal of our \textit{HST} snapshot program with WFC3 (GO 11657; PI: L. Stanghellini) was to obtain high-resolution images of a meaningful sample of angularly small PNe (i.e., with published angular diameters $<$4\arcsec). We refer the reader to SSV16 for a full description of the sample selection, observing log, goals, and implementation of the program. In this paper we summarize only the observations taken to measure the properties of the CSs
and the aspects of the sample selection that are relevant to the study of the CSs presented here. The sample includes angularly small Galactic PNe and excludes Galactic bulge and Halo PNe (see SSV16).  Since this was a snapshot program all targets have to be bright enough to be observed within one \textit{HST} orbit. Our approved program consisted of 130 targets, 42\% of the original target list was finally observed.

The observations were acquired with \textit{HST} Wide-Field Camera~3 \citep{Kimble_etal08} using four filters 
(F502N, F200LP, F350LP, and F814W).
We imaged each target with the F200LP filter, which passes light of all wavelengths to which the detectors are sensitive, and in particular takes advantage of the extraordinary sensitivity of this camera to near-UV light \citep{WFC3_ihb}. 
We complemented the NUV exposures with the F350LP filter, which blocks UV light but passes all visible light. 
The difference between the calibrated flux of the central star in F200LP and that in F350LP filters measures the UV continuum. 
Finally, we obtained an image with the F814W filter, which measures the I-band continuum.
All images except those in F814W were split into sub-exposure pairs to enable cosmic-ray rejection.  
We bracketed the exposure durations for the continuum filters to extend the dynamic range of the detections, which is necessary to account for the large range in possible absolute brightness of the CS, which can vary by 10~mag in $V$ as it evolves at nearly constant luminosity from 30,000~K to well over 100,000~K. 
For F814W we obtained a single exposure, modified for interstellar extinction. The goal was to  detect a possible companion between spectral types G2V and M5V\footnote{A significant contribution to the $V$-mag from other types of potential companions is not expected (see \citealt{VSS:04}).}. 
We used the longest exposures to obtain the highest S/N ratio unless the CS was saturated, in which case we used short exposures. {The F200LP and F350LP exposures were designed to yield a S/N of $\sim20$ for a star with apparent magnitude $V=25$ and a temperature of $10^5$~K. A second, $x$10 times shorter exposure  was also taken to avoid saturation of bright targets. All the images were processed using the standard WFC3 calibration pipeline (CALWFC3 version 2.0, 08-Mar-2010) (see SSV16). 

\subsection{Sample description}

In this paper we focus on the determination of the CS properties of the sample of  51 angularly compact PNe presented in SSV16.  The sample represent mostly a random selection (based on the sky position) out of the 130 PNe included in the original snapshot program target list which contains a large fraction of all spectroscopically-confirmed Galactic PNe whose ground-based measured radius is smaller than 4\arcsec \ in the Galaxy. The analysis of the extinction and spatial distributions presented in SSV16 shows that the sample although it belongs to a general Galactic population  it is skewed toward larger galactocentric radii populating the Galaxy outskirts. This means that this sample of compact PNe is seen at farther distances, on average, with respect to the general PN population and may be representative of brighter PNe on average.

SSV16 found that the average oxygen abundance of compact PNe
is slightly lower than that of the whole Galactic PN population.
However, the statistical sample for oxygen in compact PNe is
still too small to determine with certainly whether this is related to a Galactic chemical evolutionary effect.
Further spectroscopic ground-based
analysis of compact PNe  is in progress (Lee et al., in prep.). 
Furthermore, SSV16 also found that most of the compact Galactic
PNe  are not in an earlier dynamical evolutionary
stage than the general Galactic PN population, although
compact PNe include a large fraction of optically thick PNe.
The morphological distribution of compact PNe is similar in distribution
to that of the general Galactic sample. The sample is typical of the general Galactic population, except that our sample includes some of the physically smallest PNe, and the average distance is significantly greater. These findings concur to asses that the compact PN sample studied in this paper is unbiased towards young or massive PNe progenitors.  

\subsection{Photometric technique}

In this sample of compact PNe the nebular continuum is often comparable in surface brightness to that of of the CS, is spatially variable, and generally overlaps the CS. 
We used aperture photometry to determine the CS magnitudes, which is reasonably well suited to the high spatial resolution of our \textit{HST} images, and which typically allows the stellar and nebular flux to be spatially distinguished. 
We measured the stellar fluxes with the Aperture Photometry Tool \citep{Laher12}, using a small circular aperture of radius 5.0 pixels (0.2$\arcsec$). 
The local background was estimated with a circular annulus, with inner and outer radii that were customized for each nebula to be as large as possible while sampling the nearby nebular background where the surface brightness matched that underlying the star. 
In the end, most of the inner radii of the background annuli were near the boundary of the source aperture, and the outer radii were typically 5.0 to 10.0 pixels larger. 
Using small, proximate background areas compromises somewhat the accuracy of the stellar photometry, in that a small amount of light from a PSF extends beyond our stellar aperture. 
In these cases, however, the background is dominated by the surrounding nebula rather than the wings of the PSF. 
The background value was determined from the median value of the pixels in the annulus. 
The uncertainty in the stellar flux includes contributions from both the Poisson uncertainty in the measurement and the RMS deviation of the background about the median value. 
These uncertainties are reflected in the error associated with the magnitudes of each target. 

We transformed the measured magnitudes to aperture-corrected instrumental magnitudes (STmags) using the relations developed for STIS and WFC3 for infinite aperture, and the relation for the encircled energy as a function of radius \citep{WFC3_ihb}. 
The results for the three filters and the errors are presented in Table~\ref{Table ST magnitudes}. 

\begin{table*}
\centering
\captionsetup{justification=centering}
\caption{Magnitudes and errors. \label{Table ST magnitudes}}
\begin{tabular}{l r r r c c}
\hline\hline
PN name & F200LP$^a$ & F350LP$^a$ & F814W$^a$ & m$_{V}$$^b$ & m$_{I}$$^b$ \\ 
\hline

PN~G000.8$-07.6$ & 18.98 $\pm$ 0.05  & 19.04 $\pm$ 0.05  & 20.38 $\pm$ 0.10  & 18.90 $\pm$ 0.05  & 20.32 $\pm$ 0.10 \\
PN~G014.0$-05.5$ & 17.23 $\pm$ 0.01  & 17.44 $\pm$ 0.01  & 18.17 $\pm$ 0.03  & 17.28 $\pm$ 0.01  & 18.13 $\pm$ 0.03 \\
PN~G014.3$-05.5^1$ & \ldots & \ldots &  \ldots  & \ldots  &  \ldots \\
PN~G021.1$-05.9$ & 21.02 $\pm$ 0.10  & 21.13 $\pm$ 0.11  & 21.91 $\pm$ 0.19  & 20.98 $\pm$ 0.11  & 21.86 $\pm$ 0.19 \\
PN~G025.3$-04.6$ & 17.04 $\pm$ 0.09  & 17.00 $\pm$ 0.08  & 18.50 $\pm$ 0.09  & 16.82 $\pm$ 0.08  & 18.44 $\pm$ 0.09 \\
PN~G026.5$-03.0$ & 16.13 $\pm$ 0.01  & 16.32 $\pm$ 0.01  & 16.99 $\pm$ 0.02  & 16.15 $\pm$ 0.01  & 16.94 $\pm$ 0.02 \\
PN~G042.9$-06.9$ & 15.33 $\pm$ 0.11  & 15.30 $\pm$ 0.13  & 16.96 $\pm$ 0.11  & 15.14 $\pm$ 0.13  & 16.90 $\pm$ 0.11 \\
PN~G044.1$+05.8$ & 16.77 $\pm$ 0.01  & 16.97 $\pm$ 0.01  & 17.70 $\pm$ 0.03  & 16.89 $\pm$ 0.01  & 17.66 $\pm$ 0.03 \\
PN~G048.5$+04.2$ & 20.60 $\pm$ 0.06  & 20.71 $\pm$ 0.06  & 21.34 $\pm$ 0.14  & 20.65 $\pm$ 0.06  & 21.31 $\pm$ 0.14 \\
PN~G052.9$-02.7$ & 18.54 $\pm$ 0.02  & 18.55 $\pm$ 0.02  & 18.82 $\pm$ 0.04  & 19.06 $\pm$ 0.03  & 18.84 $\pm$ 0.04 \\
PN~G053.3$+24.0$ & 16.67 $\pm$ 0.01  & 17.47 $\pm$ 0.01  & 19.11 $\pm$ 0.05  & 17.37 $\pm$ 0.02  & 19.02 $\pm$ 0.05 \\
PN~G059.9$+02.0$ & 17.56 $\pm$ 0.02  & 17.30 $\pm$ 0.01  & 16.82 $\pm$ 0.02  & 17.89 $\pm$ 0.03  & 16.83 $\pm$ 0.02 \\
PN~G063.8$-03.3^2$ & 18.60 $\pm$ 0.03  & 18.49 $\pm$ 0.03  & 18.19 $\pm$ 0.03  & \ldots & \ldots \\
PN~G068.7$+01.9$ & 19.26 $\pm$ 0.03  & 19.21 $\pm$ 0.03  & 19.22 $\pm$ 0.05  & 19.15 $\pm$ 0.03  & 19.18 $\pm$ 0.05 \\
PN~G068.7$+14.8$ & 15.94 $\pm$ 0.03  & 16.24 $\pm$ 0.04  & 17.47 $\pm$ 0.04  & 16.10 $\pm$ 0.04  & 17.43 $\pm$ 0.04 \\
PN~G079.9$+06.4$ & 20.77 $\pm$ 0.06  & 20.81 $\pm$ 0.07  & 21.16 $\pm$ 0.13  & 20.69 $\pm$ 0.06  & 21.12 $\pm$ 0.13 \\
PN~G097.6$-02.4$ & 18.47 $\pm$ 0.02  & 18.57 $\pm$ 0.02  & 19.02 $\pm$ 0.05  & 18.39 $\pm$ 0.02  & 18.97 $\pm$ 0.05 \\
PN~G098.2$+04.9$ & 21.01 $\pm$ 0.22  & 20.88 $\pm$ 0.23  & 20.63 $\pm$ 0.20  & 21.13 $\pm$ 0.23  & 20.62 $\pm$ 0.20 \\
PN~G104.1$+01.0$ & 20.28 $\pm$ 0.07  & 20.19 $\pm$ 0.07  & 19.91 $\pm$ 0.08  & 20.71 $\pm$ 0.08  & 19.93 $\pm$ 0.08 \\
PN~G107.4$-02.6$ & 20.33 $\pm$ 0.05  & 20.35 $\pm$ 0.05  & 20.55 $\pm$ 0.09  & 20.22 $\pm$ 0.05  & 20.51 $\pm$ 0.09 \\
PN~G184.0$-02.1$ & 16.41 $\pm$ 0.01  & 16.38 $\pm$ 0.01  & 16.52 $\pm$ 0.02  & 16.27 $\pm$ 0.01  & 16.48 $\pm$ 0.02 \\
PN~G205.8$-26.7^2$ & 15.40 $\pm$ 0.01  & 16.19 $\pm$ 0.02  & 17.58 $\pm$ 0.03  & \ldots & \ldots \\
PN~G263.0$-05.5$ & 16.18 $\pm$ 0.01  & 16.13 $\pm$ 0.01  & 15.78 $\pm$ 0.01  & 15.96 $\pm$ 0.01  & 15.73 $\pm$ 0.01 \\
PN~G264.4$-12.7$ & 14.53 $\pm$ 0.01  & 15.01 $\pm$ 0.01  & 16.09 $\pm$ 0.01  & 14.87 $\pm$ 0.01  & 16.03 $\pm$ 0.01 \\
PN~G275.3$-04.7$ & 18.90 $\pm$ 0.04  & 19.43 $\pm$ 0.07  & 20.33 $\pm$ 0.09  & 19.26 $\pm$ 0.07  & 20.28 $\pm$ 0.09 \\
PN~G278.6$-06.7$ & 18.13 $\pm$ 0.11  & 19.12 $\pm$ 0.26  & 19.65 $\pm$ 0.11  & 18.98 $\pm$ 0.26  & 19.59 $\pm$ 0.11 \\
PN~G285.4$+01.5$ & 19.45 $\pm$ 0.09  & 19.38 $\pm$ 0.09  & 19.18 $\pm$ 0.07  & 19.38 $\pm$ 0.09  & 19.15 $\pm$ 0.07 \\
PN~G285.4$+02.2$ & 20.80 $\pm$ 0.07  & 20.91 $\pm$ 0.07  & 21.45 $\pm$ 0.15  & 20.76 $\pm$ 0.07  & 21.40 $\pm$ 0.15 \\
PN~G286.0$-06.5$ & 16.12 $\pm$ 0.06  & 16.29 $\pm$ 0.07  & 17.54 $\pm$ 0.06  & 16.10 $\pm$ 0.07  & 17.49 $\pm$ 0.06 \\
PN~G289.8$+07.7$ & 18.75 $\pm$ 0.03  & 19.41 $\pm$ 0.05  & 20.82 $\pm$ 0.11  & 19.30 $\pm$ 0.05  & 20.76 $\pm$ 0.11 \\
PN~G294.9$-04.3$ & 15.53 $\pm$ 0.01  & 15.69 $\pm$ 0.01  & 16.17 $\pm$ 0.01  & 15.60 $\pm$ 0.01  & 16.13 $\pm$ 0.01 \\
PN~G295.3$-09.3$ & 15.25 $\pm$ 0.03  & 15.34 $\pm$ 0.04  & 16.88 $\pm$ 0.04  & 15.17 $\pm$ 0.04  & 16.82 $\pm$ 0.04 \\
PN~G296.3$-03.0$ & 19.21 $\pm$ 0.09  & 19.27 $\pm$ 0.10  & 19.43 $\pm$ 0.08  & 19.14 $\pm$ 0.10  & 19.39 $\pm$ 0.08 \\
PN~G309.0$+00.8$ & 18.15 $\pm$ 0.06  & 18.07 $\pm$ 0.06  & 18.05 $\pm$ 0.06  & 18.15 $\pm$ 0.06  & 18.03 $\pm$ 0.06 \\
PN~G309.5$-02.9$ & 20.62 $\pm$ 0.26  & 20.55 $\pm$ 0.26  & 23.08 $\pm$ 0.43  & 20.63 $\pm$ 0.26  & 23.06 $\pm$ 0.43 \\
PN~G324.8$-01.1$ & 19.69 $\pm$ 0.04  & 19.59 $\pm$ 0.04  & 19.05 $\pm$ 0.05  & 20.35 $\pm$ 0.05  & 19.09 $\pm$ 0.05 \\
PN~G327.1$-01.8$ & 16.73 $\pm$ 0.01  & 16.68 $\pm$ 0.01  & 16.48 $\pm$ 0.02  & 16.75 $\pm$ 0.02  & 16.45 $\pm$ 0.02 \\
PN~G327.8$-06.1$ & 15.29 $\pm$ 0.01  & 15.47 $\pm$ 0.01  & 15.98 $\pm$ 0.01  & 15.29 $\pm$ 0.01  & 15.92 $\pm$ 0.01 \\
PN~G334.8$-07.4^3$ & 13.28 $\pm$ 0.01  & 13.10 $\pm$ 0.01 & 12.85 $\pm$ 0.01 & 12.29 $\pm$ 0.01 & 12.79 $\pm$ 0.01 \\
PN~G336.9$+08.3$ & 15.77 $\pm$ 0.01  & 15.89 $\pm$ 0.01  & 16.43 $\pm$ 0.01  & 15.81 $\pm$ 0.01  & 16.39 $\pm$ 0.01 \\
PN~G340.9$-04.6$ & 19.07 $\pm$ 0.09  & 19.07 $\pm$ 0.10  & 15.39 $\pm$ 0.05  & 18.99 $\pm$ 0.10  & 15.35 $\pm$ 0.05 \\
PN~G341.5$-09.1$ & 15.19 $\pm$ 0.01  & 15.78 $\pm$ 0.01  & 16.97 $\pm$ 0.02  & 15.61 $\pm$ 0.01  & 16.91 $\pm$ 0.02 \\
PN~G343.4$+11.9$ & 16.84 $\pm$ 0.01  & 17.30 $\pm$ 0.02  & 18.45 $\pm$ 0.04  & 17.14 $\pm$ 0.02  & 18.40 $\pm$ 0.04 \\
PN~G344.2$+04.7$ & 15.93 $\pm$ 0.04  & 15.88 $\pm$ 0.04  & 16.40 $\pm$ 0.04  & 15.73 $\pm$ 0.04  & 16.36 $\pm$ 0.04 \\
PN~G344.8$+03.4$ & 18.21 $\pm$ 0.02  & 18.28 $\pm$ 0.02  & 18.69 $\pm$ 0.04  & 18.10 $\pm$ 0.02  & 18.64 $\pm$ 0.04 \\
PN~G345.0$+04.3$ &  15.34 $\pm$ 0.01 & 15.33 $\pm$ 0.01 & 15.43 $\pm$ 0.01 & 15.26 $\pm$ 0.01  & 15.39 $\pm$ 0.01 \\
PN~G348.4$-04.1$ & 17.74 $\pm$ 0.05  & 17.67 $\pm$ 0.05  & 17.28 $\pm$ 0.03 & 17.65 $\pm$ 0.05  & 17.25 $\pm$ 0.03 \\
PN~G348.8$-09.0^2$ & 15.66 $\pm$ 0.01  & 16.13 $\pm$ 0.01 & 16.80 $\pm$ 0.02  & \ldots & \ldots \\
PN~G351.3$+07.6$ & 15.94 $\pm$ 0.02  & 16.25 $\pm$ 0.02  & 17.09 $\pm$ 0.02  & 16.07 $\pm$ 0.02  & 17.04 $\pm$ 0.02 \\
PN~G356.5$+01.5$ & 19.28 $\pm$ 0.03  & 19.18 $\pm$ 0.03  & 18.77 $\pm$ 0.04  & 19.63 $\pm$ 0.04  & 18.78 $\pm$ 0.04 \\
PN~G358.6$+07.8$ & 17.09 $\pm$ 0.01  & 17.20 $\pm$ 0.01  & 17.78 $\pm$ 0.03  & 17.03 $\pm$ 0.01  & 17.73 $\pm$ 0.03 \\ \hline
\end{tabular}
\begin{tablenotes}
\item
\centering
$^a$ST system.
$^b$Johnson-Cousins system.\\
$^1$CS not detected above the nebular level.
$^2$ No extinction constant available: transformations to Johnson mags not possible.
$^3$CS saturated in all images.
\end{tablenotes}
\end{table*}

\subsection{Extinction correction}
We used the nebular Balmer decrement to derive the stellar extinction correction, following the same procedure as in \citet{VSS2003}. 
The nebular extinction constants, $c$, have been taken from the literature (\citealt{Tylenda1994}; \citealt{GH2014}),  as corrected by our absolute \textit{HST} fluxes (SSV16). 
We then derived the color excess, $E(B-V)$ from the nebular extinction constants. 
According to \citet{KL1985} the ratio of $c$ to $E(B-V)$ is nearly constant and shows little variations with the stellar temperature in the range considered here. 
We used the Galactic extinction law from \citet{SM1979} to derive $E(B-V)$ of each nebula assuming $R_V=3.1$.
Note the values of $c$ in these PNe tend to be much higher than average for Galactic PNe, with $c$ approaching 3.0 in some cases (see Table~\ref{Table initialdata}). 
A large extinction value introduces a larger correction in the transformation to Johnson $V$ magnitudes and thus possibly, but not necessarily, a larger uncertainty in the derived standard magnitude. 
The $E(B-V)$ values used are given in Table \ref{Table initialdata}.

\begin{table*}
\centering
\captionsetup{justification=centering}
\caption{\label{Table initialdata}Adopted parameters of compact planetary nebulae.$^a$}
\begin{tabular}{c c c c c c c r}
\hline\hline
PN name & R$_{phot}$  & --log F(H$\beta$)  & $c$ & E($B-V$) & \ion{He}{ii} $I$(4686) & Ref.$^b$ & Distance \\*
 &  &  &  &  &  &  & (Kpc) \\ 
\hline
PN~G000.8$-07.6$ & 1.33  & 12.68 $\pm$ 0.15  & 0.39  & 0.28  & 1.16  & (6)   & 20.12 \\
PN~G014.0$-05.5$ & 3.80  & 12.32 $\pm$ 0.15  & 1.24  & 0.88  & 29.00 & (1)   & 5.60 \\
PN~G014.3$-05.5$ & 0.50  & 12.49 $\pm$ 0.15  & 1.11  & 0.79  & 5.90  & (1)   & 20.35 \\
PN~G021.1$-05.9$ & 2.06  & 12.28 $\pm$ 0.15  & 0.54  & 0.38  & 19.50 & (1)   & 10.55 \\
PN~G025.3$-04.6$ & 0.48  & 12.44 $\pm$ 0.15  & 0.71  & 0.35  & 0.81  & (4)   & 24.26 \\
PN~G026.5$-03.0$ & 2.93  & 12.37 $\pm$ 0.15  & 0.99  & 0.70  & $<$ 3.00  & (1)   & 9.00 \\
PN~G042.9$-06.9$ & 0.48  & 11.41 $\pm$ 0.05  & 0.49  & 0.29  & 0.33  & (4)   & 10.20 \\
PN~G044.1$+05.8$ & 3.84  & 13.51 $\pm$ 0.25  & 1.44  & 1.02  & \ldots & \ldots & 8.79 \\
PN~G048.5$+04.2$ & 1.54  & 13.06 $\pm$ 0.20   & 1.66  & 1.18  & 66.00 & (1)   & 15.21 \\
PN~G052.9$-02.7$ & 3.80  & 13.30 $\pm$ 0.20   & 2.92  & 2.07  & 1.52  & (4)   & 9.59 \\
PN~G053.3$+24.0$ & 2.32  & 11.51 $\pm$ 0.05  & 0.27  & 0.19  & 23.00 & (1,3)   & 8.32 \\
PN~G059.9$+02.0$ & 1.34$^{*}$ & 13.70  $\pm$ 0.25  & 2.93  & 2.08  & \ldots & \ldots & 12.75 \\
PN~G068.7$+01.9$ & 1.76  & 13.06 $\pm$ 0.20   & 1.64  & 1.16  & \ldots & \ldots & 10.17 \\
PN~G068.7$+14.8$ & 0.59  & 11.95 $\pm$ 0.05  & 0.38  & 0.27  & 2.40  & (1)   & 21.47 \\
PN~G079.9$+06.4$ & 2.77  & 13.28 $\pm$ 0.20   & 1.41  & 1.00  & 90.30 & (4)   & 9.75 \\
PN~G097.6$-02.4$ & 2.36  & 12.59 $\pm$ 0.15  & 0.80  & 0.57  & 12.20 & (4)   & 10.25 \\
PN~G098.2$+04.9$ & 1.29  & 13.36 $\pm$ 0.20   & 2.47  & 1.75  & 36.00 & (1,2)   & 9.93 \\
PN~G104.1$+01.0$ & 0.97  & 13.58 $\pm$ 0.25  & 2.92  & 2.07  & \ldots & \ldots   & 8.26 \\
PN~G107.4$-02.6$ & 3.54  & 13.21 $\pm$ 0.20   & 1.37  & 0.97  & 84.30 & (4)   & 8.71 \\
PN~G184.0$-02.1$ & 1.03  & 12.05 $\pm$ 0.05  & 1.43  & 1.01  & $<$ 1.00  & (3)   & 7.01 \\
PN~G263.0$-05.5$ & 1.35  & 12.01 $\pm$ 0.05  & 0.85  & 0.60  & 9.20  & (1)   & 2.39 \\
PN~G264.4$-12.7$ & 1.32  & 11.35 $\pm$ 0.05  & 0.32  & 0.23  & $<$ 1.00  & (1,7)   & 10.60 \\
PN~G275.3$-04.7$ & 1.70  & 12.15 $\pm$ 0.05  & 0.80  & 0.57  & 28.00 & (1)   & 10.25 \\
PN~G278.6$-06.7$ & 1.19  & 11.55 $\pm$ 0.05  & 0.50  & 0.35  & 6.00  & (1)   & 10.57 \\
PN~G285.4$+01.5$ & 1.92  & 12.31 $\pm$ 0.15  & 1.87  & 1.33  & $<$ 1.00  & (1,7)   & 5.77 \\
PN~G285.4$+02.2$ & 2.06  & 12.85 $\pm$ 0.20   & 1.28  & 0.91  & 78.00 & (1)   & 10.14 \\
PN~G286.0$-06.5$ & 0.95  & 11.90 $\pm$ 0.05  & 0.70  & 0.50  & $<$ 1.00  & (1)   & 10.91 \\
PN~G289.8$+07.7$ & 1.29  & 12.36 $\pm$ 0.15  & 0.35  & 0.25  & 45.00 & (1)   & 12.82 \\
PN~G294.9$-04.3$ & 0.98  & 11.73 $\pm$ 0.05  & 1.53  & 1.09  & 0.14  & (6)   & 11.36 \\
PN~G295.3$-09.3$ & 0.48  & 11.94 $\pm$ 0.05  & 0.44  & 0.31  & $<$ 1.00  & (1)   & 21.25 \\
PN~G296.3$-03.0$ & 1.32  & 12.01 $\pm$ 0.05  & 1.40  & 0.99  & 19.00 & (1,7)   & 7.93 \\
PN~G309.0$+00.8$ & 1.47  & 12.40 $\pm$ 0.15  & 2.03  & 1.44  & $<$ 1.00  & (1)   & 4.80 \\
PN~G309.5$-02.9$ & 1.18  & 13.09 $\pm$ 0.20   & 2.08  & 1.48  & 20.00 & (1)   & 10.95 \\
PN~G324.8$-01.1$ & 1.94  & 13.26 $\pm$ 0.20   & 3.34  & 2.37  &  \ldots  & \ldots   & 3.82 \\
PN~G327.1$-01.8$ & 1.50  & 12.27 $\pm$ 0.15  & 1.98  & 1.40  & $<$ 3.00  & (1,5)   & 7.43 \\
PN~G327.8$-06.1$ & 1.05  & 12.10 $\pm$ 0.05  & 0.51  & 0.36  & $<$ 4.00  & (1)   & 15.13 \\
PN~G334.8$-07.4$ & 0.44  & 11.19 $\pm$ 0.05  & 0.55  & 0.39  & $<$ 2.00  & (1)   & 4.68 \\
PN~G336.9$+08.3$ & 1.77  & 12.73 $\pm$ 0.15  & 1.47  & 1.04  & $<$ 7.00  & (1)   & 9.63 \\
PN~G340.9$-04.6$ & 1.12  & 12.80 $\pm$ 0.20   & 1.57  & 1.11  & $<$ 10.00 & (1)   & 12.50 \\
PN~G341.5$-09.1$ & 1.37  & 12.30 $\pm$ 0.15  & 0.40  & 0.28  & $<$ 4.00  & (1)   & 12.17 \\
PN~G343.4$+11.9$ & 1.54  & 12.45 $\pm$ 0.15  & 0.55  & 0.39  & 19.50 & (1,7)   & 14.06 \\
PN~G344.2$+04.7$ & 0.55  & 12.07 $\pm$ 0.05  & 1.17  & 0.83  & $<$ 2.00  & (1)   & 14.05 \\
PN~G344.8$+03.4$ & 2.32  & 13.04 $\pm$ 0.20   & 0.57  & 0.40  & $<$ 10.00 & (1)   & 14.29 \\
PN~G345.0$+04.3$ & 2.19$^{*}$ & 13.25 $\pm$ 0.20 & 1.53  & 1.09  & $<$ 10.00 & (1)   & 11.29 \\
PN~G348.4$-04.1$ & 0.97  & 12.73 $\pm$ 0.15  & 1.72  & 1.22  & $<$ 10.00 & (1)   & 11.27 \\
PN~G351.3$+07.6$ & 0.53  & 12.35 $\pm$ 0.15  & 0.69  & 0.49  & $<$ 4.00  & (1)   & 23.87 \\
PN~G356.5$+01.5$ & 2.28  & 13.68 $\pm$ 0.25  & 2.77  & 1.96  & \ldots  & \ldots   & 7.04 \\
PN~G358.6$+07.8$ & 1.99  & 12.54 $\pm$ 0.15  & 1.05  & 0.74  & $<$ 3.00  & (1)   & 12.64 \\ 
\hline
\end{tabular}
\begin{tablenotes}
\item
\centering
$^a$ All parameters from SSV16 except $I$(4686). \\
$^b$ Original references of the \ion{He}{ii} fluxes.\\
$^{*}$ Photometric radius containing 95\% of the flux.
\tablebib{(1) \citet{Acker1992}; (2) \citet{Aller1987}; (3) \citet{Barker1978}; (4) \citet{GH2014}; (5) \citet{KJ1991};  (6) \citet{Lee2016} (in prep.); (7) \citet{SK1989}. }
\end{tablenotes}
\end{table*}

\subsection{Transformation to standard Johnson V, I magnitudes}\label{VJohnson}

The filters in the {\it HST} instruments do not match perfectly the bandpasses of standard photometric systems, such as Johnson-Cousins $UBVRI$, so the transformation from instrumental magnitudes to a standard system depends on the spectral energy distribution (SED) of the object. 
Although ST magnitudes allow a first characterization of the CSs, calculations such as the Zanstra temperature determination, will require
the magnitude to be converted to a standard system. 

We derived the Johnson $V$ magnitudes from the STmags in the filters F200LP and F350LP, and the Johnson $I$ magnitude from the STmags in filter F814W. 
The color needed to apply the transformation has been derived via synthetic photometry with the STSDAS SYNPHOT package using a blackbody spectrum to represent the SED of the CS. 
This procedure estimates the difference between the UVIS/WFC3 filters and the hypothetical measurement in a Johnson band. 
To derive the transformation to the Johnson $V$ we assumed that the PN CSs behave as blackbodies with a range of effective temperatures (30,000 to 300,000 K \footnote{The lower temperature limit is set to provide enough ionizing photons, and the upper limit is taken from the evolutionary tracks of \citet{VW1994}.}) as in \citet{VSS2003,VSS:04,VSS:07}. 
Given that we did not know at this stage the effective temperature of the CS, we took the median of the correction in the 30,000 to 300,000 K range (using steps of 5000~K) to derive the magnitude in $V$. 
We added quadratically the standard deviation of these corrections to the error in the magnitudes. 
We explored three different methods in order to obtain the best accuracy. 
First, we have calculated the corrections covering the above temperature range, comparing the F200LP band to the $V$ band. 
Secondly, we have performed the same action using the F350LP band against the $V$ band.
Finally, we compared the derived $V$ to the band resulting from subtracting the red contribution (F814W) to the F350LP band. 
We found that the second approach, F350LP vs.\ $V$, was less sensitive to the yet unknown stellar temperature, showing the least variance within the range considered. 
Thus, we have used this transformation to assure a good stability of our results. 
The transformation is more straightforward between the ST F814W and the $I$ magnitudes, since the passbands are very similar. 

The largest standard deviation obtained in the transformation to $V$ is 0.024 mag, and the largest value of the transformation is 0.734 mag (both values correspond to the PN~G324.8$-01.1$ which is suffering the highest extinction, $c=3.34$). 
Regarding  Johnson $I$, the highest standard deviation obtained is 0.013 mag, and the largest correction is $-0.06$ mag. 
Given the great dependency of the transformation on the extinction we have opted not to transform the instrumental magnitudes to the Johnson  system for those PNe where the extinction was unknown. 
Those are PN~G$063.8-03.3$, PN~G$205.8-26.7$, PN~G$348.8-09.0$ and have been marked in Table \ref{Table ST magnitudes}. 

The transformations between filters were re-calculated using a fixed temperature instead of a range of values once we had an estimate of the effective temperature of the star (see Sect. \ref{Zanstra}). 
With the new values of the magnitudes a new effective temperature was determined. 
We iterated this procedure until we reached convergence. 
The final STmags and errors as well as standard $V$ and $I$ magnitudes are given in Table \ref{Table ST magnitudes}. 

We compare in Figure~\ref{Mags} the derived $V$ magnitudes for our sample of compact PNe with the largest sample of $V$ band photometry of Galactic PNe from Shaw and collaborators \citep{SK85,SK1989,KSK90}. 
These samples have little overlap, and the compact sample (median $V = 17.8$) clearly contains much fainter CSs than the prior Galactic sample (median $V=15.6$). 
But as noted in SSV16, the compact sample suffers considerably more interstellar extinction on average. 
Once the effect of extinction is removed, the median $V$ magnitude of the compact PN CSs is only 0.66~mag fainter. 
It is hard to argue that these samples are very different based solely upon the brightness distribution of their central stars. 

\begin{figure}
\centering
\resizebox{\hsize}{!}{\includegraphics{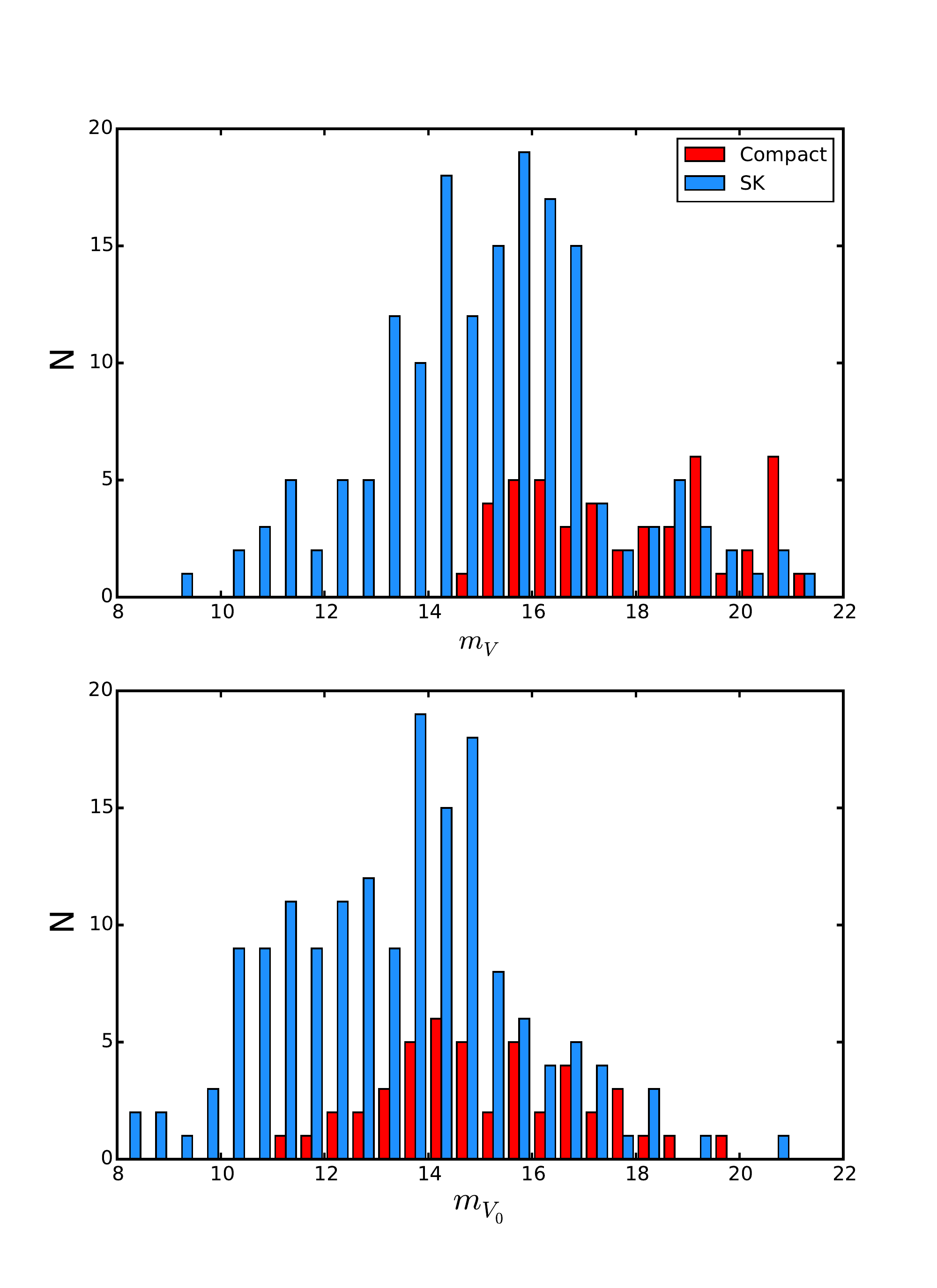}} 
\caption{Apparent central star $V$ magnitudes (\textit{upper}), and those corrected for interstellar extinction (\textit{lower}), from this sample and that from the largest Galactic sample  \citep{SK85,SK1989,KSK90} (see legend).}
 \label{Mags}
\end{figure}

\section{Determination of the CS effective temperatures and luminosities} \label{section3}
\subsection{Zanstra temperatures}\label{Zanstra}

The effective temperatures of CSs of PNe are often estimated with the technique of \citet{Zans31} as fully developed by \citet{HS1966} and \citet{Kaler83}. 
The Zanstra method derives the total ionizing flux of the star by comparing the flux of a nebular recombination line of hydrogen or helium with the stellar continuum flux in the optical band, assuming that all the photons above the Lyman limit of H ($\lambda<921$\AA) or He$^+$ ($\lambda<228$\AA) are absorbed within the nebula. At high optical depth each recombination results in a Balmer series photon. 
To apply this method we need a measure of the \ion{H}{i} and \ion{He}{ii} nebular fluxes, the nebular extinction, and the $V$ magnitude of the central star. 
While the \ion{He}{ii} $\lambda4686$ flux is typically 10\% or less of that from H$\beta$, useful measurements or upper limits are available for most of our sample. 

The presence of \ion{He}{ii} in the nebular spectrum indicates higher CS temperatures; but the presence of \ion{He}{i} also indicates the nebula is optically thick. 
Following \citet{Kaler83}, the absence of \ion{He}{i} and $I$(4686) $>90$ indicates that the nebula is optically thin even to \ion{He}{ii} ionizing radiation, so the Zanstra temperature would be a lower limit. 
However, the choice between Zanstra temperatures using \ion{H}{i} or \ion{He}{ii} is sometimes dictated by the available data. 
If $I$(4686) is not detected, which is the case for many PNe in our sample including those with upper limits, then it is unclear whether the nebula is optically thick in H continuum. 
Thus T(\ion{H}{i}) may or may not be a lower limit. 
Consequently we have adopted Zanstra temperatures derived from \ion{He}{ii} $\lambda4686$ where possible, and Zanstra temperatures using the \ion{H}{i} otherwise.
The values of H$\beta$ and $I(4686)$ used are given in Table \ref{Table initialdata}. For $I$(4686), a 10$\%$ error has been adopted. 
 
The transformation of the ST magnitudes to Johnson magnitudes provided us with correction values, which are of the order of 0.15 mag, but that can be reduced once the temperature of the CS has been determined in a first iteration. 
Thus, we have applied an iterative process using Zanstra temperatures. 
First, we have calculated the Johnson $V$ magnitudes using synthetic photometry and a range of blackbody temperatures appropriate for CSs of PNe. We then obtained a list of values for the transformation, and chosen the median value. 
Zanstra temperatures and errors were then derived using  these $V$ magnitudes. 
Once we had a first estimate of the temperature for each CS, we used it to recalculate the synthetic photometry and to obtain a new value of  the
Johnson $V$ magnitudes. 
We then used these new $V$ magnitudes to calculate again the Zanstra temperatures. 
The process was repeated until the variation in the corrections of the magnitudes (and therefore the transformations) did not change and the variation in the Zanstra temperatures obtained was negligible ($\sim100$~K). 
We have also used the final Zanstra temperatures to obtain a better adjustment for the Johnson $I$ magnitudes.
The final values for the Johnson $V$ (m$_{V}$) and $I$ magnitudes (m$_{I}$) and their corresponding errors, are given in Table \ref{Table ST magnitudes}.

\subsection{Bolometric correction}\label{BC}

To calculate the bolometric corrections (BC) of our stars we have used the calibration by \citet{VGS1996}, $BC=27.66 - 6.84 \times \log T_{eff}$, where for T$_{eff}$ we have used the \ion{He}{ii} Zanstra temperature when available.
This BC was derived by \citet{VGS1996} and \citet{Flower1996} for O and early B spectral types assuming a maximum T$_{eff}$=50,000 K. 
Our CSs can reach much higher T$_{eff}$, but given there is no other calibration for hot stars available in the literature, we have assumed the relation to be valid for our targets. 
We note that, as discussed in \citet{VGS1996}, this relation depends only weakly on the surface gravity of the star.

\subsection{Luminosities}

Absolute luminosities have been calculated from our magnitudes, bolometric corrections, and distances. 
The statistical distances are those of SSV16, which used the calibration by Stanghellini et al. (2008). 
We adopted a global 30\% as the minimum distance uncertainty, attributable to the calibration of the distance scale itself \citep{SSV2008}. 
First we transformed the apparent $V$ magnitudes (see Sect. \ref{VJohnson}) to absolute $V$ magnitudes by means of the distance-magnitude relationship. 
Adding the bolometric correction (obtained in Sect. \ref{BC}), we derived the absolute bolometric magnitudes. 
The iteration methodology used when deriving Zanstra temperatures reduced the value of the magnitude error to the order of the instrumental error. 
Thus the errors in the luminosities are dominated by the large errors in the adopted distances.
In Table \ref{Table luminosity} we present the derived values for CS temperatures, the ratio (TR) of these temperatures, luminosities and bolometric corrections. 

\begin{table*}
\centering
\captionsetup{justification=centering}
\caption{ Derived central star properties.\label{Table luminosity}}
\begin{tabular}{c r r c r r}
\hline\hline
PNG name & $T_{H~I} $ & $T_{He~II} $ & TR & $log(L_{\ast}/L_\odot)$  & BC \\
 & $(10^{3}K)$ & $(10^{3}K)$ &  &  &  \\ 
\hline
PN~G000.8$-07.6$ & 47.8$\pm3.10$  & 60.2$\pm1.39$   & 1.26   & 2.96$\pm0.27$ & $-5.03\pm0.16$  \\
PN~G014.0$-05.5$ & 42.9$\pm2.55$  & 83.1$\pm2.73$   & 1.94   & 2.88$\pm0.28$ & $-5.99\pm0.23$  \\
PN~G021.1$-05.9$ & 137.2$\pm15.19$ & 130.0$\pm6.76$ & 0.95   & 2.48$\pm0.30$ & $-7.32\pm0.36$ \\
PN~G025.3$-04.6$ & 34.8$\pm1.80$  & 52.2$\pm1.10$   & 1.50   & 3.79$\pm0.27$ & $-4.61\pm0.14$  \\
PN~G026.5$-03.0$ & 32.2$\pm1.55$  & \ldots          & \ldots & 2.62$\pm0.29$ & $-3.17\pm0.33$  \\
PN~G042.9$-06.9$ & 41.7$\pm0.98$  & 51.4$\pm0.61$   & 1.23$^a$ & 3.69$\pm0.27$ & $-4.57\pm0.08$  \\
PN~G044.1$+05.8$ & 23.2$\pm1.70$  & \ldots          & \ldots & 1.91$\pm0.33$ & $-2.20\pm0.50$  \\
PN~G048.5$+04.2$ & 68.2$\pm11.05$ & 119.9$\pm10.33$ & 1.76   & 2.84$\pm0.35$ & $-7.08\pm0.59$  \\
PN~G052.9$-02.7$ & 41.9$\pm5.00$  & 60.2$\pm2.68$   & 1.44   & 2.26$\pm0.29$ & $-5.03\pm0.30$ \\
PN~G053.3$+24.0$ & 71.1$\pm0.61$  & 100.3$\pm1.60$  & 1.41   & 3.42$\pm0.26$ & $-6.55\pm0.11$  \\
PN~G059.9$+02.0$ & 27.2$\pm2.30$  & \ldots          & \ldots & 2.03$\pm0.35$ & $-2.67\pm0.58$  \\
PN~G068.7$+01.9$ & 44.6$\pm5.55$  & \ldots          & \ldots & 1.91$\pm0.43$ & $-4.14\pm0.85$  \\
PN~G068.7$+14.8$ & 37.9$\pm0.23$  & 59.2$\pm0.46$   & 1.56   & 4.12$\pm0.26$ & $-4.98\pm0.05$  \\
PN~G079.9$+06.4$ & 56.9$\pm8.35$  & $>105.1^b$  & 2.01   & $>2.04^b$ & $<-6.37$  \\
PN~G097.6$-02.4$ & 46.1$\pm2.90$  & 76.7$\pm2.35$   & 1.66   & 2.87$\pm0.27$ & $-5.75\pm0.21$  \\
PN~G098.2$+04.9$ & 68.3$\pm11.89$ & 109.4$\pm9.30$  & 1.60   & 2.17$\pm0.36$ & $-6.81\pm0.58$  \\
PN~G104.1$+01.0$ & 53.5$\pm7.50$  & \ldots          & \ldots$^a$ & 1.33$\pm0.46$ & $-4.68\pm0.96$  \\
PN~G107.4$-02.6$ & 52.2$\pm7.20$  & 107.8$\pm8.45$  & 2.07   & 2.40$\pm0.34$ & $-6.76\pm0.54$  \\
PN~G184.0$-02.1$ & 40.3$\pm0.22$  & \ldots          & \ldots$^a$ & 2.62$\pm0.26$ & $-3.84\pm0.04$  \\
PN~G263.0$-05.5$ & 36.8$\pm0.22$  & 67.2$\pm0.74$   & 1.83$^a$ & 2.42$\pm0.26$ & $-5.36\pm0.07$  \\
PN~G264.4$-12.7$ & 40.1$\pm0.26$  & \ldots          & \ldots & 3.54$\pm0.26$ & $-3.83\pm0.04$  \\
PN~G275.3$-04.7$ & 83.3$\pm0.77$  & 112.0$\pm1.95$  & 1.34   & 2.97$\pm0.27$ & $-6.88\pm0.12$  \\
PN~G278.6$-06.7$ & 127.1$\pm10.33$ & 102.5$\pm3.44$ & 0.81   & 3.01$\pm0.29$ & $-6.61\pm0.23$  \\
PN~G285.4$+01.5$ & 85.7$\pm7.61$  & \ldots          & \ldots$^a$ & 2.11$\pm0.36$ & $-6.08\pm0.61$  \\
PN~G285.4$+02.2$ & 80.8$\pm14.15$ & 134.4$\pm12.74$ & 1.66   & 2.58$\pm0.37$ & $-7.42\pm0.65$  \\
PN~G286.0$-06.5$ & 39.9$\pm0.22$  & \ldots          & \ldots$^a$ & 3.06$\pm0.26$ & $-3.81\pm0.04$  \\
PN~G289.8$+07.7$ & 67.3$\pm5.36$  & 109.1$\pm4.59$  & 1.62   & 3.12$\pm0.29$ & $-6.80\pm0.29$  \\
PN~G294.9$-04.3$ & 41.8$\pm0.27$  & 48.5$\pm0.37$   & 1.16   & 3.53$\pm0.26$ & $-4.39\pm0.05$  \\
PN~G295.3$-09.3$ & 31.6$\pm0.16$  & \ldots          & \ldots$^a$ & 3.74$\pm0.26$ & $-3.12\pm0.03$  \\
PN~G296.3$-03.0$ & 97.4$\pm3.71$  & 113.3$\pm2.73$  & 1.16$^a$ & 2.81$\pm0.27$ & $-6.91\pm0.16$  \\
PN~G309.0$+00.8$ & 54.8$\pm3.85$  & \ldots          & \ldots$^a$ & 1.91$\pm0.32$ & $-4.75\pm0.48$  \\
PN~G309.5$-02.9$ & 69.3$\pm12.15$ & 99.6$\pm7.65$   & 1.44   & 2.34$\pm0.35$ & $-6.53\pm0.53$  \\
PN~G324.8$-01.1$ & 63.4$\pm9.85$  & \ldots          & \ldots$^a$ & 1.00$\pm0.50$ & $-5.19\pm1.06$  \\
PN~G327.1$-01.8$ & 41.4$\pm2.40$  & \ldots          & \ldots$^a$ & 2.52$\pm0.31$ & $-3.92\pm0.40$  \\
PN~G327.8$-06.1$ & 30.2$\pm0.17$  & \ldots          & \ldots & 3.34$\pm0.26$ & $-2.98\pm0.04$  \\
PN~G334.8$-07.4$ & 29.7$\pm0.11$  & \ldots          & \ldots$^a$ & 3.25$\pm0.26$ & $-2.93\pm0.03$  \\
PN~G336.9$+08.3$ & 26.6$\pm1.10$  & \ldots          & \ldots & 2.59$\pm0.28$ & $-2.61\pm0.28$  \\
PN~G340.9$-04.6$ & 50.3$\pm6.93$  & \ldots          & \ldots & 2.30$\pm0.46$ & $-4.50\pm0.94$  \\
PN~G341.5$-09.1$ & 29.0$\pm1.25$  & \ldots          & \ldots & 2.97$\pm0.29$ & $-2.86\pm0.30$  \\
PN~G343.4$+11.9$ & 36.6$\pm1.95$  & 72.6$\pm2.0$    & 1.98   & 3.58$\pm0.27$ & $-5.59\pm0.19$  \\
PN~G344.2$+04.7$ & 34.8$\pm0.21$  & \ldots          & \ldots$^a$ & 3.27$\pm0.26$ & $-3.40\pm0.04$  \\
PN~G344.8$+03.4$ & 33.0$\pm3.30$  & \ldots          & \ldots & 2.27$\pm0.38$ & $-3.25\pm0.68$  \\
PN~G345.0$+04.3$ & 29.2$\pm0.13$  & \ldots          & \ldots & 3.06$\pm0.29$ & $-2.88\pm0.30$  \\
PN~G348.4$-04.1$ & 38.3$\pm2.10$  & \ldots          & \ldots$^a$ & 2.42$\pm0.30$ & $-3.69\pm0.38$  \\
PN~G351.3$+07.6$ & 31.4$\pm1.45$  & \ldots          & \ldots & 3.47$\pm0.29$ & $-3.10\pm0.32$  \\
PN~G356.5$+01.5$ & 38.0$\pm4.20$  & \ldots          & \ldots & 1.21$\pm0.40$ & $-3.67\pm0.76$  \\
PN~G358.6$+07.8$ & 35.4$\pm1.85$  & \ldots          & \ldots & 2.68$\pm0.30$ & $-3.46\pm0.36$  \\ 
\hline
\end{tabular}
\begin{tablenotes}
\item
\centering
$^a$May be optically thick in H Lyman based on surface brightness. \\
$^b$I(4686)$>90$ suggests nebula is optically thin in He$^{+}$, so T(\ion{He}{ii}) and L(\ion{He}{ii}) are lower limits. \\
\end{tablenotes}
\end{table*}

\section{Results and discussion} \label{section4}

In Figure~\ref{HR-H Diagram} we show the location of the CSs in the log~$L$--log~$T$ diagram together with the post-AGB evolution tracks of \citet{VW1994} for single stars with solar metallicity.  
Both H-burning CS and He-burning CS are represented. 
Next to each track we give the main-sequence and the post-AGB CS mass; the filled squares are CSs for which T(\ion{He}{ii}) is available; empty circles are stars with T(\ion{H}{i}) only, and may be lower limits. 
\textbf{}

It is well known that Zanstra temperatures obtained from \ion{He}{ii} are higher than (and more closely approximate the true effective temperature) those obtained with \ion{H}{i}. 
This has generally been interpreted as an optical depth effect \citep[the so-called Zanstra discrepancy; see][]{Kaler83}. 
We find the same behavior in our sample, with only two exceptions, PN~G021.1$-05.9$ and PN~G278.6$-06.7$ (see Table~\ref{Table luminosity}). 

\citet{Vmg:02} showed that the transition from an optically thick to an optically thin nebula depends on the initial mass of the star: the higher the initial mass, the higher the effective temperature at which the nebula becomes optically thin. 
If the Zanstra discrepancy is due only to the optical thickness in the H-ionizing radiation and since TR approaches unity for higher effective temperatures, then according to the results of \cite{Vmg:02}, it is very likely that the objects with the higher TR have lower-mass progenitors. 
The fact that  PN~G021.1$-5.9$, the CS, shows the smallest Zanstra discrepancy and also it turns out to have one of the largest progenitor mass of our sample (see Table \ref{Table masses}) corroborates this idea. We note that the mass could not be determined for PN~G278.6$-06$. 
Furthermore, all the objects in our sample with the TR larger than two have low mass CSs and the most massive CS PN~G68.7+14.8 has a very small TR. Note that the morphology plays a role in the thin-thick nature of the nebula given that a bipolar object can be optically thick in the torus but thin in the lobes. However, as pointed out by \cite{Vmg:02} observations are dominated by the brightest shell of the nebula and this is usually the optically thick internal region able to trap the ionization front.

From the location of the stars in the HR diagram and their effective temperatures we can easily argue that we are not dealing with very young PNe. This is specially true for CSs with initial masses M $<2$ M$_{\odot}$ that require a few thousand years to leave the constant luminosity part of the HR diagram. 
Thus the compact nature of most of the PN must be due to their larger distances. 
The average galactocentric distance of compact PNe is $10.10\pm5.01$ \,kpc, a Galactic sample of hundreds of objects analyzed by \cite{SH10} has an average of $6.56\pm4.14$ \,Kpc. If we also compare them with the  mean distance, $2.87$ kpc, of the all PNe analyzed by \cite{Manchado2004} (angularly resolved nebula observed from the ground from the \cite{Manchado1996} catalog) for which a distances estimate is available we see that our sources are located a factor of four farther away on average. 
So the small angular size of many of the sources in our sample must be attributed to large distances, rather than to their possible younger nebular ages. In SSV16, the average properties of the sample presented here have been analyzed against the Galactic PNe sample of \cite{SH10} confirming that on average compact Galactic PNe are farther away from the Galactic center than the general population. 

\begin{figure} 
\centering
 \resizebox{\hsize}{!}{\includegraphics{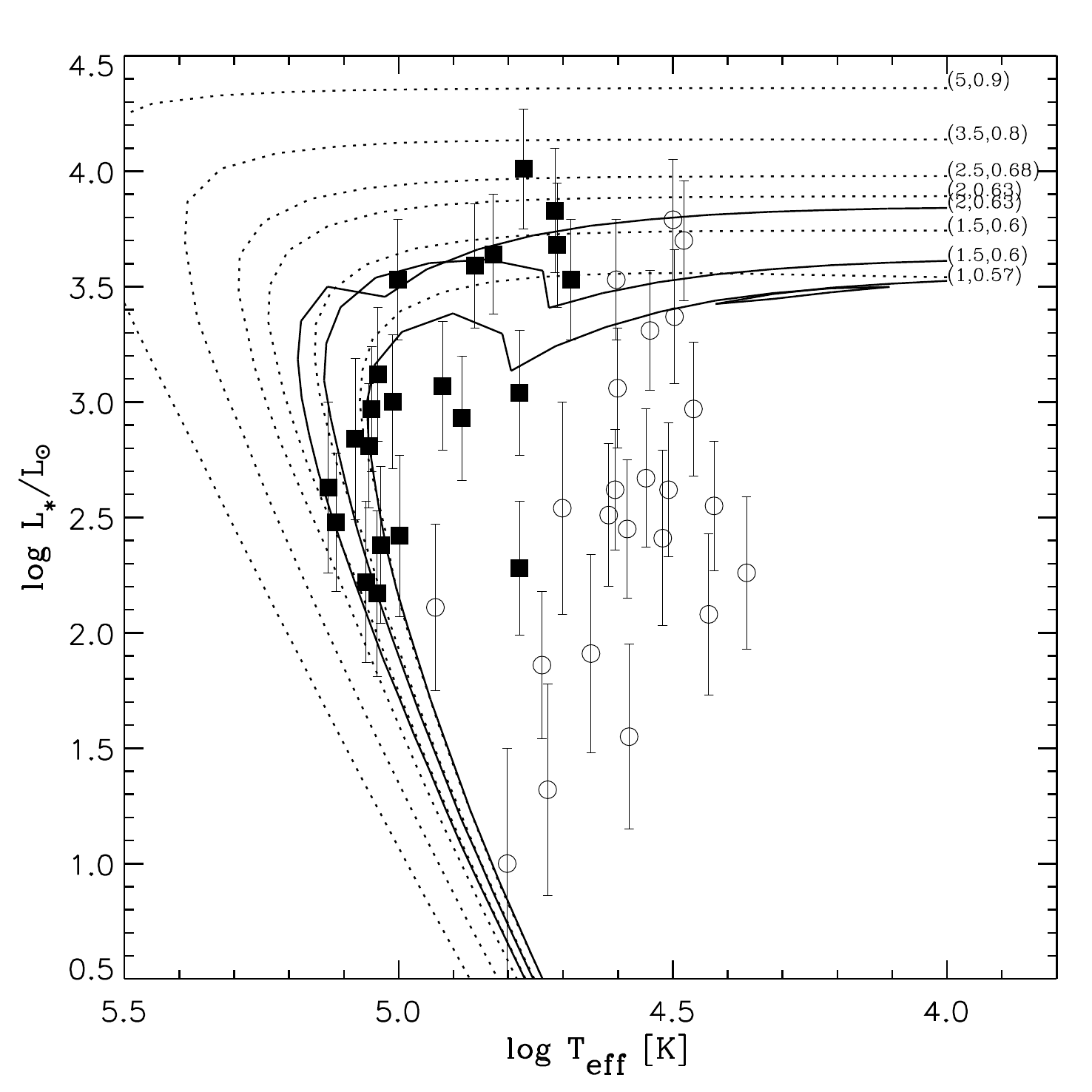}}
\caption{HR diagram showing the location of the CSs from this sample. 
Filled squares represent CSs with T$_{eff}$ determined from \ion{He}{ii} and the empty circles indicate T(\ion{H}{i}). 
Curves representing H- (dashed) and He-burning (solid) tracks for Z=0.016 from \citet{VW1994} are also plotted, and labelled with the corresponding MS and post-AGB masses.} 
 \label{HR-H Diagram}
\end{figure}

The determination of the CS masses has been done according to their location in the log~$L-$log~$T$ plane. 
If the star is located in the constant-luminosity part we can derive the masses directly from the relation given by \citet{VW1994}, $L/L_{\odot}=56694 \cdot \left(\frac{M}{M_{\odot}}-0.5\right) \label{Masseq}$. 
Otherwise we have interpolated from the value obtained by using the nearest tracks to the CS location. 
We have assumed an error of 0.025 \Mso\ when the CS mass is determined in this way. 
Finally, if the CS location falls too far below of any of the evolutionary tracks available we were not able to determine the mass, this is the case for 24 of the objects analyzed (see  Fig.~\ref{HR-H Diagram}). 
We believe that some CS luminosities may be inaccurate because of erroneous distances, and that some temperatures adopted from T(\ion{H}{i}) may in fact be lower limits. 
To obtain the CS masses from the He-burning tracks we have used the same general procedure described above. 
For those CSs located in the bump region of the HR diagram we have also interpolated between the closest tracks to obtain the value for its mass. 
We assumed again an error of 0.025~\Mso. %
The inferred CS masses are listed in Table \ref{Table masses}. We have several  CSs with locations under the lowest mass track (either H or He tracks) but close enough to them that we have determined their masses. Those objects are prepended with a colon in in Table \ref{Table masses}; these objects are easy to identify in Figure~\ref{HR-H Diagram}.

Given the large number of objects that fall outside the theoretical evolutionary tracks on the HR diagram we have revisited the issue of the uncertainties we carried out in the CS parameter determination. We identified two major sources of error (besides the instrumental ones) that we can quantify: the effective temperature of the CS, and the distance to the PN. Looking into the effective temperatures it is important to note that all the objects, with one exception, for which we cannot derive accurate masses have hydrogen Zanstra effective temperatures determinations and that for many of these PNe only upper limits to their 4686 \AA~\ion{He}{ii} fluxes from ground-based spectra are available (although in many cases it can establish hard limits to the temperature). Thus we can argue that the temperatures derived for these objects are just lower limits and then explore what would be the increase on temperature required for these CSs to cover the evolutionary tracks. If we consider only the temperature we need that the bulk of the objects move from effective temperatures of $\approx 40,000$ \,K to twice this value. Given that the well known Zanstra discrepancy usually involves a factor of two in the derivation of the temperature this does not seem to be an unreasonable scenario. Furthermore,  an increase in the CS temperature not only implies moving the CS on the horizontal axis on the HR diagram but also in the vertical direction since it forces a change in the magnitude of the bolometric correction. A change of the CS temperature from $36,000$ to $72,000$ \,K moves it  $0.7$ \,dex in luminosity ($1.1$ \,dex if we allow the temperature to increase from $36,000$ to $100,000$ \,K). Thus increasing a factor of two the CS temperature of the objects with hydrogen Zanstra values moves their location in the HR diagram to be within the tracks. 

Let us now revisit the issue of the distance uncertainty in our CS mass determinations. The distance scale we used is that derived by \cite{SSV2008}, which is an improved calibration of the \cite{CKS} statistical distance using Magellanic Cloud objects. The distances are determined  if the angular diameter, the 5 GHz flux, and the optical thickness parameter $\tau$ are available (see SSV16). Note that the apparent radii were not available for these compact PNe before the {\it HST} images and thus this is the first reliable determination of their statistical distances. The photometric radii are shown in Table \ref{Table initialdata}. The photometric radius gives an objective measurement of nebular angular size that is insensitive to the S/N ratio of the image and is useful for evolution studies. It is derived according to the method described by \citet{Stanghellini1999}. It corresponds to the size of a circular aperture that contains 85\% of the flux in [\ion{O}{iii}] $\lambda$5007.

The applied distance scale assumed an uncertainty of at least 30\% for any PN due to the uncertainty in the available PN ionized masses. The parameter $\tau$ is used to discriminate between optical thin and thick nebulae: for $\tau<$2.1 the PN is considered optically thick, and its ionized mass depends on the progression of the ionization front toward the outer nebula; while for $\tau>$2.1 the PNe is optically thin and the ionized mass is assumed constant. It is beyond the scope of this work to grasp whether the distance scale works better for optically thick nebula than for optically thin objects for which a constant ionized mass is assumed. It is worth to note, however, that in SSV16 we found that for compact PNe $<\tau>$=2.7$\pm$0.85, higher on average than the general PN Galactic sample. It is even more relevant to recall that out of all the PNe for which we have hydrogen Zanstra temperatures and fall outside the tracks, more than half of them have $\tau> 2.1$. 

Errors in the distance will affect every object in a different way. However, let us explore the possibility that a systematic error  is affecting the luminosity and causing so many CSs to fall outside the theoretical evolutionary tracks. The distance changes the location of all objects (those with hydrogen and Helium Zanstra effective temperatures) in the vertical axis. If distances are underestimated by a factor of two all CSs locations will move in the HR diagram to higher luminosities by  $1.24 dex$. This will put half of the objects that now fall outside within the theoretical tracks. However this change has two important consequences: i) the average CS mass of the sample moves from 0.59 $M_{\odot}$ to 0.7 $M_{\odot}$ which is much higher than any other CS mass average found so far in either Galactic or extragalactic environments and ii) this change in the distance will put a large fraction of the sample population outside of the nominal value given for the Galaxy limits which would be quite unrealistic. We cannot exclude the possibility that a underestimation of the distance could be the cause behind so many stars falling outside the tracks. However,  having such a high average CS mass (0.7$M_{\odot}$) and large average distances ($\approx$20 kpc) 
would point to a very unusual selection bias of the sample and given that we have not found that our PNe sample of compact objects is different in any significant way to a general Galactic sample having such a high average CS mass it does not seem reasonable to assume that we are dealing with this scenario.

On the other hand having the distance overestimated by a factor of two the CSs move downwards on the HR leading to an older population and to the impossibility of deriving masses for a larger number of CSs. So although we certainly cannot exclude a systematic error in the distance determination we think it seems more reasonable to assume that the hydrogen Zanstra temperatures derived for our sample of objects are underestimated at least a factor of two. This is certainly a well documented behavior in the literature. Furthermore, the fact that most of the objects with hydrogen Zanstra determinations also seem to be optically thin points in the same direction and possibly (although we have not provided probe of this) to larger errors in their distances. 

If some of the CSs have unresolved companions they could contribute significantly to the flux, though that contribution would be larger in the I band if the companion is cool. Such a contribution represents a potentially large source of systematic error in the derived temperatures and luminosities of the CSs. A companion is certainly a possibility for PN~G044.1+5.8, where SSV16 commented on the very bright central star embedded in a faint, morphologically mature nebula. Indeed, the derived L and T are well within the lowest mass evolution track. To quantify this possibility in depth would require exploring contributions from all potential classes of companions, including main sequence stars, red dwarfs, white dwarfs, and Giants, as well as modeling the nebular continuum. Narrowing the range of possibilities further would require follow-up spectroscopy on the central star. These investigations are beyond the scope of this paper, but merit a comprehensive follow-up study.

The He- or H-burning nature of the post-AGB track is determined by whether helium shell or hydrogen shell--burning is dominant when the star leaves the AGB. 
It has been shown (i.e, \citealt{Iben1984, Wood1986, Schoenberner1986, Renzini1989}, and references therein), that the evolutionary path leading to a He-shell burning CS is uncommon, with less than 25\% of all the stars that enter the PN phase ending up as He-burners. 
On the other hand, the Vassiliadis \& Wood (1994) models are more efficient at producing He-burning post-AGB tracks for low mass progenitors, which they argue is a natural consequence of the mass-loss behavior during the AGB phase. 
Since the mechanism that controls the departure of the star from the AGB is unknown and therefore artificially defined in the stellar evolutionary models we have preferably assumed H-burning tracks to estimate the mass of our CSs. 
The differences in the CS mass inferred using He- or H-burning tracks are negligible for most of the stars in our sample and they are inconsequential when compared with other sources of error in the mass determination. 
The initial metallicity of the star is expected to play a role into the initial-to final-mass relation. A lower metallicity environment would result for the same initial mass progenitor into a higher CS core mass, with caveats, since it has been shown that mass-loss can have a dependency on the carbon- or oxygen rich nature of the sources. Measurements of dust and gas mass-loss rates in the Magellanic Clouds and dwarf irregular galaxies show no decrease in mass-loss rates for the carbon-stars. This is because of these stars produce all the carbon they need to drive the mass loss rates (see e.g., \citealt{Mat07, Sloan09}). For oxygen-rich sources there is indeed an observed decrease in their mass-loss rates with metallicity because low- and intermediate-mass stars do not produce oxygen during their lifetimes (i.e., \citealt{Sloan10}).
We have no a priory reason to assume that the initial composition of the progenitors stars in our sample was lower than the general assumed for the Galaxy. However given the large distances and the fact that some of them could be located in the outer parts of the Galaxy we can assume the metallicity of the LMC and see how the average mass determined changes under this assumption using the Vassiliadis \& Wood (1994) tracks. The CS mass for those objects located in the horizontal part of the HR diagram has an analytical expression and it does not depend on the metallicity.  Thus the mass determined for the CSs does not change with the metallicity involved. The dependency with the metallicity is in the initial-final mass relation. Since we do not provide the initial mass values, the metallicity assumed for the sample is not relevant in the discussion of our results. 

\begin{table}
\centering
\captionsetup{justification=centering}
\caption{PN core mass.\label{Table masses}}
\begin{tabular}{l c c}
\hline\hline
PNG name & M/\Mso & M/\Mso \\
 & H track & He track \\ 
 \hline
PN~G021.1$-05.9$ &         0.615 & 0.617  \\
PN~G025.3$-04.6$ &         0.619 &  $:$0.634    \\
PN~G042.9$-06.9$ &         0.586 & 0.617  \\
PN~G048.5$+04.2$ &         0.583 & 0.584  \\
PN~G053.3$+24.0$ &         0.569 & 0.601  \\
PN~G068.7$+14.8$ &         0.732 & \ldots  \\
PN~G079.9$+06.4$ &         0.597 & 0.600  \\
PN~G098.2$+04.9$ &         0.597 & 0.600    \\
PN~G107.4$-02.6$ &         0.569 & 0.567  \\
PN~G264.4$-12.7$ &         0.569 & 0.614  \\
PN~G275.3$-04.7$ &         $:$0.569 & 0.567  \\
PN~G285.4$+01.5$ &         $:$0.569 & $:$0.567  \\
PN~G285.4$+02.2$ &         0.633 & 0.634  \\
PN~G289.8$+07.7$ &         $:$0.569 & 0.567  \\
PN~G294.9$-04.3$ &         0.569 & 0.617  \\
PN~G295.3$-09.3$ &         0.596 & 0.596  \\
PN~G296.3$-03.0$ &         0.569 & 0.567  \\
PN~G309.5$-02.9$ &         0.569 & 0.567  \\
PN~G324.8$-01.1$ &         $:$0.569 & $:$0.567  \\
PN~G327.8$-06.1$ &         \ldots & $:$0.567  \\
PN~G343.4$+11.9$ &         0.583 & 0.600    \\
PN~G344.2$+04.7$ &         \ldots & $:$0.567  \\
PN~G351.3$+07.6$ &         0.552 & 0.584  \\ 
\hline
\end{tabular}
\begin{tablenotes}
\item
\centering
The symbol $:$ means the CS is under the lowest mass track\\ 
but close enough that we can determine its mass. \\
\end{tablenotes}
\end{table}

Out of the 51 objects analyzed we were able to determine CS masses for 23. 
Of these, we searched for correlations between CS mass and nebular macro-morphological type (taken from SSV16) and dust mineralogy \citep[taken from][]{Stan:12}, but found none. 
This is perhaps not surprising given the small sample size and, perhaps more importantly, the narrow range of CS masses. 
Most of the CS masses are relatively small, with only one CS exceeding 0.65~$M_{\odot}$ corresponding to a progenitor mass $>2$ $M_{\odot}$.
  
\begin{figure}
\centering
\resizebox{\hsize}{!}{\includegraphics{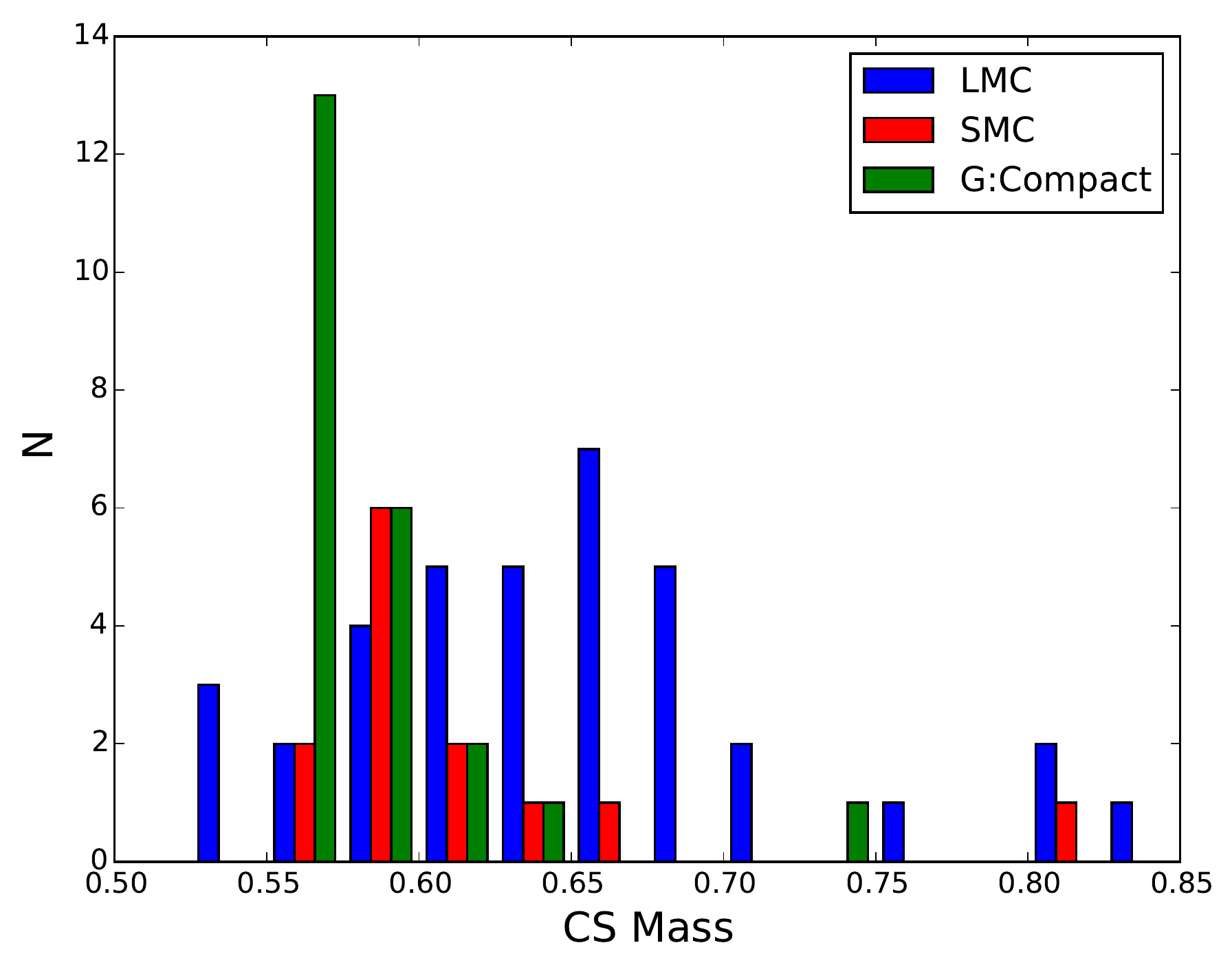}} 
\caption{Central star mass distributions for the PNe analyzed by: \cite{VSS:03,VSS:07} in the LMC (\textit{blue}), \cite{VSS:04} in the SMC (\textit{red}) and the Galactic sample analyzed here (\textit{green}). 
Masses have been derived using a mixture of the H- and He-burning tracks from \cite{VW1994}, though this choice makes little difference with a bin size of 0.025~\Mso.}
 \label{Morph}
\end{figure}

The most massive star in our sample has 0.73~\Mso\ and the least massive has 0.55~\Mso. 
The mean of the sample is 0.59~\Mso\ with a spread of 0.04~\Mso\ using the H-burning tracks; if He-burning tracks are used the mean is 0.56~\Mso\ with a slightly smaller spread (because there are fewer objects) of 0.02\Mso. 
These average masses are consistent with the mass distribution for Galactic white dwarfs 0.6~\Mso$\;$\citep{Weidemann1983, Weidemann1990} and with white dwarf mass determinations from the Sloan Digital Sky Survey 0.59~\Mso\ \citep{Hu2007}. \cite{SVMG2002} also
obtained a similar value (0.6~\Mso) from the analysis of $\sim$200 northern Galactic CSs of PNe. 

In Figure~\ref{Morph} we show the CS mass distribution histogram 
compared to the mass distribution histograms previously obtained for the LMC \citep{VSS:03,VSS:07} and the SMC \citep{VSS:04}.  
Although dealing with a small sample of objects that have been selected in a particular way (their angular size), we compare them with similar size samples obtained with the {\it HST} and analyzed using the same technique: PNe in the Magellanic Clouds \cite{VSS2003, VSS:04, VSS:07}. 
The mean CS mass obtained for the Large Magellanic Cloud PNe is 0.65~\Mso\ \citep{VSS:07} which is significantly higher than that our compact Galactic PNe. 
The Small Magellanic Cloud CS sample \citep{VSS:04} also has a slightly higher mean, which was shown to have a mean mass of 0.63~\Mso. 
The Large and Small Magellanic Clouds have, on average, lower metallicity ($\sim$1/2 and $\sim$1/4 respectively) than the Galaxy \citep{RB1989, RD1990}. 
Metallicity has a strong influence on the mass-loss during the AGB phase, which is driven mostly by dust \citep{Wood1979,Bowen1988}. 
Thus, low-metallicity stars with dust-driven winds are expected to lose smaller amounts of matter \citep{Willson2000}. 
Consequently, it is to be expected that the CSs of PNe mass distribution showed greater mean values for the LMC and SMC than for the Galaxy. 
Although we used a small sample of stars, we see this tendency in our  sample of compact PNe observed with {\it HST} in the Galaxy.
It is also important to point out that available accurate mass measurements of the most abundant  white dwarfs in the Galaxy (those with hydrogen atmospheres) strongly peak at 0.6~\Mso (see i.e., \citealt{Kepler}).

\section{Conclusions} \label{section5}

We have analyzed 51 compact PNe imaged with \textit{HST} with the goal of disentangling the CS from the nebula in order to determine the masses of the CSs for the first time using direct measurements of the CS magnitude. 
We determined magnitudes, CS temperatures and luminosities for most stars in the sample. 
From the location of the CSs on the H--R diagram, and by comparison with the evolutionary tracks of \citet{VW1994} we inferred the masses of 23 objects. 
The average mass of this small sample of stars is 0.59~\Mso, which does not significantly change whether we use H- or He-burning tracks. 
The average masses of the CSs in our sample agree with the masses found in larger Galactic samples. 

Finally, in agreement with the conclusions of SSV16, we attribute the compact nature of most sources in this sample to their larger average distances rather than their evolutionary status. 
CS mass determinations of PNe are subject to large distance uncertainties. 
This and other limitations of our method prevented the determination of the mass for 24 of the stars in our sample. 
However, the anticipated GAIA data releases\footnote{See http://www.cosmos.esa.int/web/gaia/release} \citep{Caci} will allow a re-determination of the CS positions in the HR with accurate distances using the measurements we are providing. 
We will then be able to test whether the misplacement of the CS relative to the tracks is due to the distance or to other still yet to be explored problems in the mass determination. 
In any event, this sample of PNe probes much deeper into the Galactic disk, and offers the prospect of probing a mix of distant stellar populations, Galactic abundance gradients, and other subjects for which distant objects offer an advantage. 

\begin{acknowledgements}
Partial support for this work was provided by NASA through grant GO--11657 from Space Telescope Science Institute, which is operated by
the Association of Universities for Research in Astronomy, Incorporated, under NASA contract NAS5--26555. 
We thank support from the Spanish Ministerio de Economia y Competitividad under project AYA2014-55840P.
\end{acknowledgements} 

\bibliographystyle{aa}
\bibliography{bibliografia}

\end{document}